\documentclass{iopart}  
\usepackage{iopams}  
\usepackage[utf8]{inputenc}
\usepackage{hyperref}
\hypersetup{colorlinks=true,allcolors=blue}
\usepackage{hypcap}
\usepackage{array}
\usepackage{graphicx}
\graphicspath{{images/},{figures/}}
\usepackage[english]{babel}
\usepackage{lineno}

\begin{document}

\newcommand{\RSM}{\text{RSM}}
\newcommand{\ext}{{\rm ext}}

\title{Quantum measurement optimization by decomposition of  measurements into extremals}

\author{Esteban Mart\'inez-Vargas$^{1}$, Carlos Pineda$^2$, Pablo Barberis-Blostein$^3$}

\newcommand{\cu}{Universidad Nacional Aut\'onoma de M\'exico, M\'exico D. F. 01000, Mexico}
\address{$^1$
Física Teòrica: Informació i Fenòmens Quàntics, Departament de Física, Universitat Autònoma de Barcelona, 08193 Bellaterra (Barcelona), Spain}
\address{$^2$Instituto de Física, Universidad Nacional Autónoma de México, \cu}
\address{$^3$Instituto de Investigaciones en Matem\'aticas Aplicadas y Sistemas, \cu}
\begin{abstract} 
  Using the convex structure of positive operator value measurements
  and of several quantities used in quantum metrology, such as quantum
  Fisher information or the quantum Van Trees information, we present
  an efficient numerical method to find the best strategy allowed by
  quantum mechanics to estimate a parameter. This method explores
  extremal measurements thus providing a significant advantage over
  previously used methods. We exemplify the method for different cost
  functions in a qubit and in a harmonic oscillator and find a strong
  numerical advantage when the desired target error is sufficiently
  small.
%
%
%
%
%
\end{abstract} 

\maketitle

\tableofcontents
\section{Introduction}
The goal of quantum metrology is to find limits in the precision of
parameter estimation of quantum
systems~\cite{escher_quantum_2011,fujiwara_strong_2006,Giovannetti2011,
  ConservativeMankei2017,true_precision2015Jarcyna}. Recent research
includes theoretical and experimental
advances~\cite{Geza2014QuantumMetr,Colangelo2017}. Knowing these limits
allows to know if a given measurement strategy minimizes the
estimation errors. The strategy that minimizes the error will be
called the optimal measurement strategy. Given a quantum state that
depends on a set of parameters, the optimal measurement strategy for
estimating these parameters can be used to estimate the quantum
state. These strategies has applications in quantum technologies, for
example they can can be used together with quantum control for quantum
state manipulation~\cite{O'Brien2009}.

To estimate a parameter of a physical setup one acquires data through
measurements and the estimation of the parameter is obtained by applying
a function, known as the estimator, to the data. 
Data has a random component. The probability distribution of measurement outcomes can be
modelled using a statistical model of the experiment: the probability
distribution of outcomes conditioned to the value of the parameter.
Given a mathematical model of a system using quantum mechanics, the
statistical model is obtained once it is decided which operator is
going to be measured. Notice that we can include classical noise in 
the quantum mechanical description. Going from the data to the estimation of the
parameters that minimizes the error is not trivial.

Given a cost function that quantifies the error, the optimal measurement strategy consist of
the quantum measurement and estimator that extremizes it. But, finding
the extreme of a cost function over all possible quantum measurements
and estimators is not simple. When a Cramér-Rao type inequality
exists, the problem can be reduced to find the extreme of another cost
function over all the quantum measurements. This simplifies the
problem because it is not longer necessary to maximize over the space
of estimators.

It is possible, though very costly, to numerically find the maximum
over all quantum measurements of cost functions. The straightforward
way is to randomly sample the space of all positive operator value
measures (POVMs), evaluate the cost function on this sample, and keep
the maximum value obtained. This method, that we will call the random
sampling method (RSM), is very inefficient because the POVM space is
large.

In this paper we show how using the algorithm proposed by Sentís et
al. \cite{sentis_decomposition_2013}  we can numerically
find the maximum over all quantum measurements of some cost
functions, that are useful in quantum metrology, in a way that is
orders of magnitude faster than using the RSM. 
This paper presents a direct application of this algorithm. We used it as
a method to extract extremal POVMs from a general POVM. It is easy to produce
efficiently a general POVM from a random unitary matrix, however, it is not trivial
to produce randomly extremal POVMs, this is where the algorithm plays an essential role.

We will call our method the random extreme sampling method (RESM). The techniques here
presented can be used, for example, to find numerically the quantum
van Trees information ~\cite{PhysRevA.95.012136} or the quantum Fisher
information in the case that the initial state is not pure. Together
with the value of the cost function maximum, the quantum measurement
that maximizes it is obtained.

We start presenting the mathematical tools, including the 
Crámer-Rao inequality and its quantum extensions in 
\sref{sec:mathematical:tools}. In the following section we 
present the RESM in detail (\sref{sec:convexnum}). We finish 
the bulk of this article 
comparing the performance of RESM against RSM in a two-level scenario, and
benchmarking the accuracy of the method against other
ansatz for a harmonic oscillator in two different
physical situations. 
%
%
We close
the article with some concluding remarks in \sref{sec:conclusions}.


\section{Mathematical tools}
\label{sec:mathematical:tools}
We discuss how to get the best parameter estimation given a statistical
model. Then we discuss how to apply this ideas for a quantum system.
\subsection{Crámer-Rao inequality}
\label{sec:cramer:rao:naked}
Now we introduce some basic quantities needed to develop further discussion.
Let 
\begin{equation}
p(\mathbf{y}|\theta)
\label{eq:distribution:probability}
\end{equation}
be the distribution
probability of the
outcomes $\mathbf{y}$, of the random variable $\mathbf{Y}$,
conditioned to a fixed value of the real parameter $\theta$. We assume
that each $\mathbf{y}$ is a set of real numbers of fixed finite size. This is the
statistical model. The function $\hat{\theta}(\mathbf{y})$ is called the
estimator and gives and estimation of
$\theta$. The estimator is unbiased when is in average correct,
 \begin{equation}
   \langle\hat{\theta}(\mathbf{y})\rangle=\int \rmd\mathbf{y}\,
   p(\mathbf{y}|\theta)\hat{\theta}(\mathbf{y})=\theta\, .
\label{eq:unbiased}
 \end{equation}
The uncertainty of the estimator is given by the mean squared error 
\begin{equation}
   \varsigma^2\equiv
       \int(\hat{\theta}(\mathbf{y})-\theta)^2p(\mathbf{y}|\theta)\rmd\mathbf{y}\, .
\end{equation}
We say that the measurement strategy is optimal if the estimator
minimizes the uncertainty.
Finally, let us define the Fisher information
\begin{equation}
F(\theta)\equiv
   \int\left(\frac{\partial\ln p(\mathbf{y}|\theta)}{\partial\theta}\right)^2
                p(\mathbf{y}|\theta)\rmd\mathbf{y}.
\end{equation}

If the estimator is unbiased (i.e. if \eref{eq:unbiased} holds), using 
the Cauchy-Schwartz inequality, one arrives to the Cram\'er-Rao inequality
\cite{cramer_mathematical_1945,Rao1992}:
\begin{equation}
	\varsigma^2 F(\theta)\geqslant 1.
	\label{cramerrao}
\end{equation}
Note
that $\varsigma^2$ depends on the choice of the specific estimator
($\hat{\theta}(\mathbf{y})$), whereas the Fisher information depends
only on the probability density function of the random variable.
From the Cram\'er-Rao inequality, we see that the inverse of the Fisher information
bounds from below the mean squared error independently of the
estimator we use; the larger the Fisher information is the smaller
the error bound. The best measurement strategy saturates
the Cram\'er-Rao inequality. Fisher showed that in the limit
where the number of measurements goes to infinity, the maximum
likelihood estimator saturates this inequality \cite{fisher_mathematical_1922}.
\subsection{Bayesian Cramér-Rao inequality}

In general each time a measurement is done, the parameter is different. 
Let us give two examples. The first is a length that changes due to 
thermal fluctuations. The second, a parameter (such as
the phase of a phase shifter) characterizing an element of 
an ensemble of non-identical objects. 
Both situations can be modelled assuming that the parameter to be measured
is a random variable (with its characteristic distribution). 
%
A third case, in which a random variable approach is suitable, is 
when the parameter is fixed, but we have some partial knowledge 
contained in an {\it a priori} distribution. 
The Bayesian
Cramér-Rao inequality can be used to decide what is the best estimator
in this situation. 

Lets assume now that each time we make an experiment the parameter we
want to estimate is different. We model the parameter as the random
variable $\Theta$, with outcomes $\theta$, and probability
distribution $\lambda(\theta)$. The outcomes of the experiment are
modelled as the random variable $\mathbf{Y}$, with outcomes
$\mathbf{y}$, and probability distribution $p(\mathbf{y}|\theta)$. The
experiment is modelled in the following way: first we take a value,
$\theta_1$, from the random variable $\Theta$; the outcome of the
experiment is $\mathbf{y}_1$ which is taken from the random variable
$\mathbf{Y}$ with distribution probability $p(\mathbf{y}|\theta_1)$;
using $\mathbf{y}_1$ we can estimate $\theta_1$ by
$\hat{\theta}(\mathbf{y}_1)$,
After repeating the
experiment $n$ times, we have $n$ estimations $\hat{\theta}_i$ with
$i=1\dots n$. The error of experiment $i$ is
$(\hat{\theta}_i-\theta_i)$. The mean squared error is, after
performing the experiment $n$ times,
$(1/n)\sum_{i=1}^n(\hat{\theta}(\mathbf{y}_i)-\theta_i)^2$. This is the cost
function we want to minimize over all the estimators $\hat{\theta}$.
In the limit $n\rightarrow\infty$ it can be written as
\begin{equation}
  \Xi^2=\int(\hat{\theta}(\mathbf{y})-\theta)^2P(\mathbf{y},\theta)\rmd\theta
  \rmd\mathbf{y}\, ,
\end{equation}
where $P(\mathbf{y},\theta)=p(\mathbf{y}|\theta)\lambda(\theta)$. It
can be shown that the error is bound from below by the Cram\'er-Rao
type inequality\cite{trees_detection_2004},
\begin{equation}
  \Xi^2Z\geqslant1\, ,
  \label{vantreescramerao}
\end{equation}
where the generalized Fisher information, $Z$, can be written as
\begin{equation}
  Z=\int F(\theta)\lambda(\theta)\rmd\theta +
  \int\left(\frac{\partial\ln\lambda(\theta)}{\partial\theta}\right)^2\lambda(\theta)\rmd\theta\, .
  \label{vantreeslambda}
\end{equation}
The first term of the sum is the expectation value of the Fisher
information, the second term is the Fisher information of the
probability distribution of the possible values of the parameter. Note
that we already know something about the parameter; this a priory
knowledge is given by $\lambda(\theta)$. As can be seen from the
previous equation, the generalized Fisher information is larger than
the Fisher information of the knowledge we already have of the
parameter. This has a simple interpretation: we can use
$\lambda(\theta)$ to estimate the parameter and measuring the system
necessarily  diminishes the error in the estimation of the parameter.
The best strategy for measuring the outcomes of a random variable is
given by the estimator, $\hat{\theta}$, that saturates this
inequality.


In this context, we found useful~\cite{RevModPhys.83.943}, a review of bayesian
inference in physics. 
\subsection{Quantum Cramér-Rao inequality}
\label{sec:QuFiIn}

We want to find the best measurement strategy to estimate a parameter,
$\theta$, that appears in the Hamiltonian of a quantum system. In
order to estimate the parameter we proceed as follows: we start with
an initial state and let the system evolve some time, after which the
state of the system is $\rho(\theta)$. One then measures some
observable of the system; we use the result to estimate $\theta$.
Fixing the Hamiltonian, time and initial state, we want to know if the
strategy we are using minimizes the error in the parameter estimation.

Measurements in quantum mechanics are described by the 
positive operator valued measure (POVM), which we briefly 
recall in order to fix the notation. 
%
If $\{E(\xi)\}$ is a POVM
parametrized by the real parameter $\xi$, for each value of $\xi$,
$E(\xi)$ is a self-adjoint operator on the system Hilbert space, they
satisfy
\begin{equation}
  \int\hat{E}(\xi)\rmd\xi=1\, ,
\end{equation}
and the probability of measuring the result $\xi$ is
\begin{equation}
  \label{eq:povm_dist}
  p(\xi|\theta)=Tr(\rho(\theta)E(\xi))\, .
\end{equation}
Notice that $\xi$ can also belong to a finite set (or a combination of several
discrete and continuous indices), if the number of possible
outcomes is finite. The expressions throughout this article generalize 
replacing $\int \rmd\xi$ by $\sum_\xi$. 

Fixing the POVM, and thinking of 
\eref{eq:povm_dist} as the distribution probability of the outcomes [as in 
\eref{eq:distribution:probability}], one 
can use the tools introduced in \sref{sec:cramer:rao:naked}. In particular,
we can calculate the Fisher information and use the 
Cramér-Rao
inequality to know if a given estimator is optimal.
Note that there exist
a dependence of the Fisher information on the POVM
we choose. In order to have the lowest bound for the error we maximize
the Fisher information over all the possible
measurements~\cite{braunstein_statistical_1994}
\begin{equation}
  \label{eq:quantum_fisher}
  F_Q(\theta)=\max_{\{\hat{E}(\xi)\}}\int\left(\frac{\partial\ln p(\xi|\theta)}
    {\partial\theta}\right)^2p(\xi|\theta)\rmd\xi.
\end{equation}
The quantity $F_Q(\theta)$ is known as the quantum Fisher information
and through the Cram\'er-Rao inequality,
\begin{equation}\label{eq:cr_quantum}
  \varsigma^2F_Q(\theta)\geqslant1\, ,
\end{equation}
tells us the minimal possible error for the best measurement strategy
for estimating a parameter appearing in the Hamiltonian of a quantum
system. \Eref{eq:cr_quantum} holds since Cram\'er-Rao inequality is valid for
every POVM, therefore it is valid for the one in which the maximum Fisher
information is attained. The POVM that maximizes $F_Q$ is the one that
should be used to get the smallest error in the parameter
estimation~\cite{escher_quantum_2011}; we call this POVM the optimal POVM. If
the
quantum state is pure there are analytical formulas for finding
$F_Q$, otherwise no general formulae are know and one must rely in numerical 
methods.
\subsection{Bayesian quantum Crámer-Rao inequality}
\label{sc:vantrees_quantum}
Assume now that each time we prepare the quantum system we are going
to meaure, the parameter $\theta$ we want to estimate is taken from a
random variable $\Theta$ with probability distribution
$\lambda(\theta)$.

The POVM, $E_{\rm max}(\xi)$, that maximizes the generalized Fisher
information,
$\int\left(\frac{\partial\ln
        P(\mathbf{y},\theta)}{\partial\theta}\right)^2P(\mathbf{y},\theta)\rmd\theta
    \rmd\mathbf{y}$,
  together with the appropiate estimator, saturates the Cram\'er-Rao
  type inequality
\begin{equation}
  \label{eq:cr_qvt_nuestra} \varsigma^2\widetilde{Z}_Q\geqslant1\, ,
\end{equation}
where
\begin{equation}
  \widetilde{Z}_Q=\max_{\{\hat{E}(\xi)\}}\left(\int\left(\frac{\partial\ln
        P(\mathbf{y},\theta)}{\partial\theta}\right)^2P(\mathbf{y},\theta)\rmd\theta
    \rmd\mathbf{y}\right)\, .
  \label{ec:ourvantrees}
\end{equation} 
We call $\widetilde{Z}_Q$ the quantum Van Trees information.

If we want to minimize the error in the parameter estimation, and we
codify what we know about the parameter in the probability
distribution $\lambda(\theta)$, we have to implement the quantum
measurement given by $E_{\rm max}(\xi)$ \cite{PhysRevA.95.012136}.

\section{Numerical calculations}
\label{sec:convexnum}
The calculation of cost functions as $F_Q$ or $\widetilde{Z}_Q$ is not
easy, as it implies an optimization over all POVMs. In this section we
present an efficient numerical procedure to calculate the maxima, over
all POVMS, of convex cost functions.
\subsection{Convexity}

%


The quantum Van Trees information is
convex; this follows directly noticing that set of
POVMs~\cite{0305-4470-38-26-010,Haapasalo2012} and the Fisher
information are convex.
Fisher information can be rewritten as $F = \int (p')^2/p \rmd\mathbf{x}$, where
the prime indicates derivative with respect to $\theta$. It then follows
that
\begin{eqnarray}
\frac{1}{2}\int\frac{(p_1^{\prime})^2}{p_1}\rmd\mathbf{x}
    + \frac{1}{2}&\int\frac{(p_2^{\prime})^2}{p_2}\rmd\mathbf{x}
    -\int\frac{[(p_1^{\prime}+ p_2^{\prime})/2]^2}{(p_1+p_2)/2} \rmd\mathbf{x} \nonumber \\
& = \frac{1}{2}\int\frac{1}{p_1p_2(p_1+p_2)}
     \left[p_1^{\prime}p_2-p_1p_2^{\prime}\right]^2\rmd\mathbf{x}
\ge0,
\end{eqnarray}
provided that $p_{1,2}\ge 0$ \cite{leyvrazprivate}. For a combination with
other weights, 
continuity and a recursive procedure imply convexity of $F$.  
But then, the Van Trees information is also convex, as the integral of 
convex functions is also convex. 

Since the maximum of the convex cost functions lies on the extremal
points of all POVMs, we only need to search in this subset simplifying
greatly the optimization task. A way to sample randomly such a set is
presented in the following paragraphs.

\subsection{The algorithm} 
The outline of the algorithm is as follows: We produce a random POVM,
and decompose it in extremals. We then evaluate the cost function
using all its extremal POVMs and choose the one which reaches the
highest value. We repeat the procedure several times and keep the
optimal POVM. We provide an implementation in \cite{repositorymetrology}.

\paragraph{Random POVMs} 
\label{sec:randpovm}

To produce a random POVM, we use backwards the purification
algorithm~\cite{nielsen2000quantum} which transforms a general POVM into a usual
projective measurement in an enlarged space.  We start from a unitary matrix
$U$ acting on a Hilbert space resulting of the product of the original Hilbert
space, and an ancilla space of dimension equal to the number of
outcomes of our POVM.  This matrix is chosen according to the Haar measure in
the enlarged Hilbert space~\cite{mehta}. To generate a member
from such ensemble we build a matrix $A$ with Gaussian complex numbers (with
equal standard deviation and zero mean). The desired matrix, $U$, is one that
diagonalizes  $A+A^\dagger$. Each of the operators $Q_m$ of the
POVM we want to generate are defined via its matrix elements as
\begin{equation}
\langle i|Q_m|j\rangle = U_{im;j1}
\label{eq:building:povm}
\end{equation}
where we are using tensor index notation for the space in which $U$
acts, the first corresponding to the original space, and the second to
the additional ancilla space. Since for all POVMs
one can build a unitary transformation in the extended space such that
\eref{eq:building:povm} holds~\cite{nielsen2000quantum}, sampling all
unitaries in the extended space guaranties sampling all POVMs with the
corresponding number of outcomes.
\paragraph{Conversion to a rank-1 POVM} 
To proceed further, we need a rank-1 POVM, so we must transform the
aforementioned POVM accordingly. Recall that a rank-1 POVM is one whose
elements are all rank-1 operators. For all operators of the POVM that are not
rank-1, we apply standard procedures to decompose them, for example the
spectral decomposition for normal operators.
%
%
Notice that the number of elements can change after this
step. Let us call $n$ the number of outcomes of the rank-1 POVM obtain.

%

\paragraph{Obtaining an element of the decomposition} 

Let us define $a_i = \tr Q_i$, and $A_{ij} = a_j^{-1} \tr (Q_j G_i)$ with 
$\{G_j\}$ an orthonormal traceless base for hermitian matrices of the 
appropriate dimension. In our case, we used the Gell-Mann matrices. We also 
define $A_{d^2,j}=1$, so that the completeness condition over POVMS
reads
$$
Aa=b,
$$
if we define the $d^2$-dimensional vector $b=(0,\cdots,0,d)$. We now propose the 
linear program
\begin{equation}
\setlength\arraycolsep{1.5pt}
  \begin{array}{l@{\quad} r c r }
    \mathrm{find}          & x &  &    \\
    \mathrm{subject~to}     &   Ax & = & b, \quad x \ge 0.   
  \end{array}
\end{equation}
Even though $x=a$ is a solution to the problem, the usual numerical algorithms 
provide an extremal point, which defines an extremal
POVM~\cite{Haapasalo2012,sentis_decomposition_2013}. Notice that if an element
$x_i$ of the solution is 0, it means that we do not include the operator 
in the POVM.
Let this extremal solution be $x^\ext$.

To obtain the extremal POVM we start by defining $x'$ via 
\begin{equation}
a = p x^\ext + (1-p)x',
\end{equation}
with $p$ a scalar.  Requiring that $x'\ge 0$ can be enforced letting 
\begin{equation}
p = \min_i \frac{a_i}{x_i^\ext}, 
\end{equation}
which in turn implies that $p$ is a probability and that for some $i$,
$x'_i=0$. If we define $\textbf{Q}^\ext = x^\ext\textbf{Q} $ and $\textbf{Q}' =
x'\textbf{Q} $, we can write
$$
\textbf{Q} = p \textbf{Q}^\ext + (1-p) \textbf{Q}'.
$$

Indeed, $\textbf{Q}$ is an extremal POVM~\cite{sentis_decomposition_2013}, and 
since one of the elements of $x'$ is null, $\textbf{Q}'$ is a $n-1$ output POVM
for which we can iterate the algorithm until a single output POVM is obtained.

Notice that with this algorithm, all POVMs with a given number of outputs
can in principle be sampled.


\section{Examples}
\label{sec:examples}
In this section we apply the method described in section \ref{sec:convexnum} to
estimate the quantum Fisher information and the
quantum Van Trees information, $\widetilde{Z}_Q$. We observe a big advantage in
terms of numerical effort using this method compared with finding the
maximum using randomly chosen POVMs.
\subsection{Qubit}
\label{sec:qubit}
We consider a spin $1/2$ particle in a superposition pure state
(see~\cite{barndorffnielsen_fisher_2000}),
\begin{equation}
  |\psi(\theta,\eta)\rangle = \left( \begin{array}{c}
    e^{-i\theta/2}\cos(\eta/2) \\
    e^{i\theta/2}\sin(\eta/2) \end{array} \right),
  \label{eq:qubits}
\end{equation}
where $\theta\in[0,2\pi)$ is the phase between the two basis states and
$\eta\in[0,\pi]$ is a known parameter that characterizes the weight of
each element of the superposition. The problem is the following: we
want to find the best estimate strategy for the phase $\theta$ when
$\eta$ is known.

Through the rest of the article we shall consider the following sets of 
POVMs:
\begin{equation}\label{eq:two_elements_povms:general}
\mathbf{P}=\{\rho(\xi),\mathcal{I}-\rho(\xi) \}_\xi\,
\end{equation}
so each POVM is parametrized by $\xi$ and has two elements that
corresponds to the outcomes $1$ and $2$. In this subsection we shall
consider the particular case
\begin{equation}\label{eq:two_elements_povms}
\rho(\xi) = |\psi(\xi,\eta)\rangle\langle\psi(\xi,\eta)|\}\,.
\end{equation}
%
%
Using
Eq.~(\ref{eq:povm_dist}) we obtain that the probability of measuring
outcome $1$ or $2$ for the POVM $\xi$ is given by
\begin{eqnarray}
  \label{eq:outcomes_qubit}
  p_\xi(1|\theta)&=&\langle\psi(\theta,\eta)|\psi(\xi,\eta)\rangle\langle\psi(\xi,\eta)|\psi(\theta,\eta)\rangle\, ,
  \nonumber\\
  p_\xi(2|\theta)&=&1-p_\xi(1|\theta)\, .
\end{eqnarray}
The  Fisher information for this probability distribution is 
\begin{equation}\label{eq:fisher_qubit_complete}
  F^{(\xi,\eta)}(\theta) =
  \frac{\sin^2(\eta)}{1+\cos^2(\eta)\tan^2((\xi-\theta)/2)}\, ,
\end{equation}
which is a function of the parameter we want to estimate, i.e. $\theta$. Notice
that there is a dependence on the initial state, via 
$\eta$,  and on the POVM used, via $\xi$. 
We make this dependence on the POVM explicit via a superscript.
When the state is pure, the maximum quantum Fisher
information can be analytically calculated \cite{braunstein_statistical_1994};
in
this example the quantum Fisher information is the maximum of
$F^{(\xi)}(\theta)$ with respect to $\xi$:
\begin{equation}
  \label{eq:quantum_fisher_qubit}
  F_Q(\theta)=\max_\xi  F^{(\xi,\eta)}(\theta)=\sin^2(\eta)\, .
\end{equation}

\begin{figure} 
    \centering
    \includegraphics{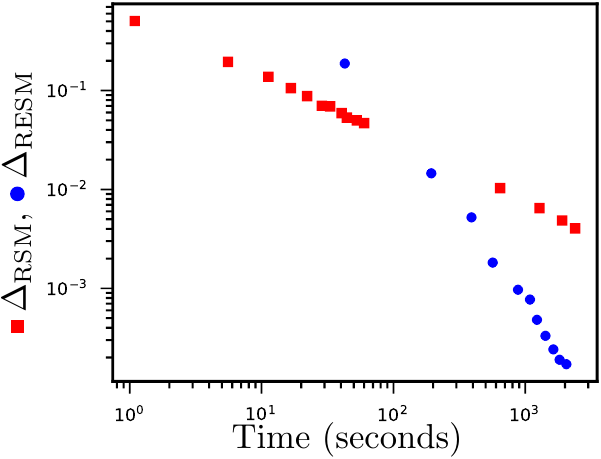}
    \includegraphics{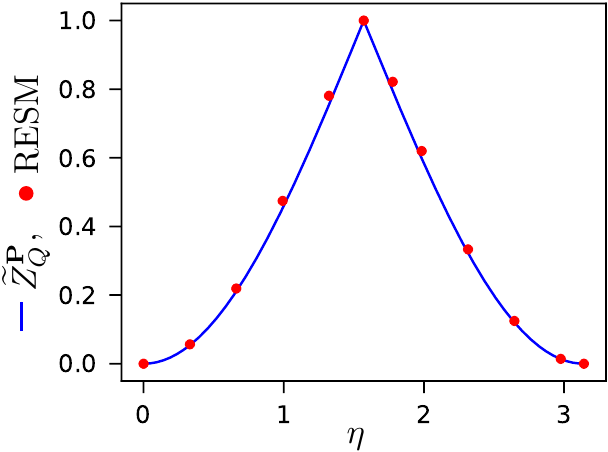}
    \caption{
    Consider estimating $\theta$ given the state \eref{eq:qubits}. 
    (left) Error in the numerical estimation of $F_Q$, see
    \eref{eq:quantum_fisher_qubit}, sampling directly the whole space of POVMs (RMS) or
    only its extremal points (RESM). We plot the error with respect to the
    computational time invested.  The slope for RSM case is $m_{\rm RSM} =
    -0.63$ and for RESM is $m_{\rm RESM} = -1.84$. The error with the method
    proposed decreases much faster using RESM than using RMS. 
      (right) 
Comparison between the numerical exploration using RESM [dots] and the ansatz 
\eref{ec:ourvantrees:maximization}, resulting in \eref{eq:zqp:our} [blue line].
Dots were numerically calculated using RESM, after sampling the space of
extremal POVMs $1000$ times. The number of outcomes of the POVM is 4, as this is
the maximum number of outcomes for an extremal POVM in a 2-dimensional Hilbert space
~\cite{0305-4470-38-26-010}.
      }
    \label{fig:tiempo_vs_error}
\end{figure}

In order to know if the RESM is useful, we apply the RSM and RESM
methods and compare their results with the exact result
Eq.~(\ref{eq:quantum_fisher_qubit}). We define the errors
\begin{eqnarray*}
  \label{eq:error_resm_qbit}
  \Delta_{\rm RSM}&=&F_Q(\theta)-F_{\rm RSM}\, ,\\
  \Delta_{\rm RESM}&=&F_Q(\theta)-F_{\rm RESM}\, ,
\end{eqnarray*}
where $F_{\rm RSM}$ and $F_{\rm RESM}$ are the Fisher information numerically
calculated using the RSM and RESM respectively. In
Fig.~\ref{fig:tiempo_vs_error}(left) we plot running time vs error, for
the two methods. It is clear from the plot that RESM is better and the
longer the program runs the better the results using RESM compared
with RSM.
For this example, we obtain an error two orders of magnitude smaller running
the program the same time.

Now we consider that $\theta$ is a random variable with probability
distribution $p(\theta)$; limits to the error in the estimation of its
outcomes are given by the Cramér-Rao type inequality
Eq.~(\ref{eq:cr_qvt_nuestra}). First we consider the maximization of
the generalized Fisher information over the family of POVMs,
$\mathbf{P}(\xi)$, given by \eref{eq:two_elements_povms:general} and 
\eref{eq:two_elements_povms}
\begin{equation}
  \widetilde{Z}_Q^{\mathbf{P}}=\max_{\xi}\int \rmd\theta p(\theta) F^{(\xi)}(\theta)\, .
  \label{ec:ourvantrees:maximization}
\end{equation} 
Because we are using a subset of all the POVMS
$\widetilde{Z}_Q^{\mathbf{P}} \leq \widetilde{Z}_Q$, nevertheless this
approach allow us to get an analytic approximation for the quantum van
Trees information. We assume that $\theta$ has a uniform distribution
in $[0,2\pi)$, i.e. $p(\theta)=1/2\pi$ in \eref{ec:ourvantrees:maximization}.
For a uniform superposition ($\eta=\pi/2$), the Fisher information
becomes independent of $\xi$; in fact $\widetilde{Z}_Q^{\mathbf{P}}
=\widetilde{Z}_Q=1$, see \eref{eq:fisher_qubit_complete}. This implies
that any POVM from the family $\mathbf{P(\xi)}$ maximizes the Fisher
information. 
%
%
In general we obtain
\begin{equation}
    \widetilde{Z}_Q^{\mathbf{P}}(\eta) = 1-|\cos(\eta)|\, ,
\label{eq:zqp:our}
\end{equation}
so we can assert that if only POVMs of the family $\mathbf{P(\xi)}$ are
allowed, the best estimation is in the case where $\eta=\pi/2$.

Now we apply RESM to calculate $\widetilde{Z}_Q$ and 
compare them with $\widetilde{Z}_Q^{\mathbf{P}}(\eta)$, see
\fref{fig:tiempo_vs_error}(right). The maximum of
$\widetilde{Z}_Q$ is obtained when $\eta=\pi/2$. That means that the
lowest error in the phase estimation is obtained when the weights of
the superposition are the same. The figure suggest that
$\widetilde{Z}_Q^{\mathbf{P}} =\widetilde{Z}_Q$.
\subsection{Phase estimation}
\label{sec:phase:estimation}




We want to estimate the phase difference $\theta$ between two paths that light
can follow, see~\cite{PhysRevA.95.012136} for a similar calculation. 
We probe the system with a coherent state, such that one path yields
the state
$|\alpha\rangle$ (with $\alpha$ a complex number) and the other 
%
\begin{equation}
    |\phi(\theta)\rangle = e^{i\hat{n}\theta}|\alpha\rangle=|e^{i\theta}\alpha\rangle\, ,
\label{eq:rotated:coherent}
\end{equation}
where $\hat{n}$ is the number operator.

\begin{figure} 
    \centering
    \includegraphics{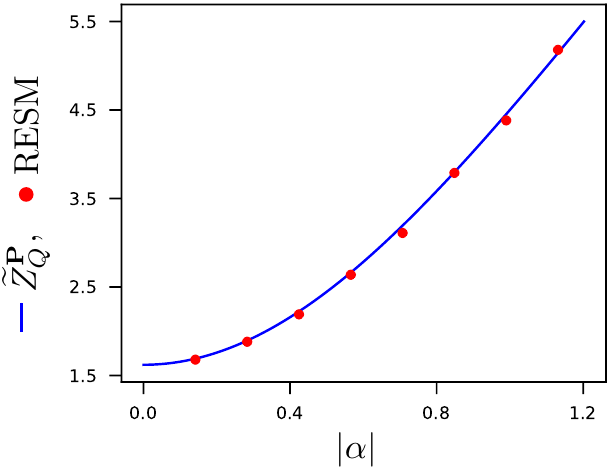}
    \includegraphics{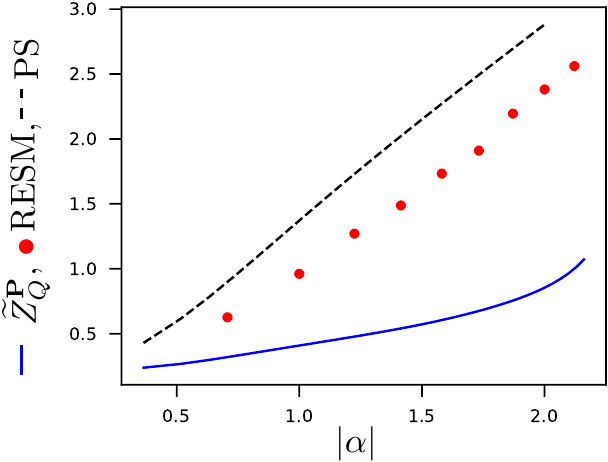}
    \caption{
    (left) Quantum van Trees information for estimating a Gaussian
       distributed random phase acquired by a coherent state.
    (right) Calculation of $\widetilde{Z}_Q$ for a equiprobable mixture of
      coherent and thermal state for a random phase chosen from Gamma
      distribution. It can be seen how the ansatz does not work and that a pure
      state (PS) is better for estimating a phase than a mixture of coherent and
      thermal state. For both figures, the dimensionless temperature is
      $k_BT=10^{-3}$, we sampled 150 times, we considered a 
      POVM with 10 outcomes and we truncate the photon space to the lowest seven 
      states of the harmonic oscillator. 
    }
    \label{fig:implaserfig}
\end{figure}

Assume that the object that creates the phase difference is subject
to fluctuations such that the phase difference between the two paths
is different each time the experiment is done. One can model this assuming
that $\theta$ is a random variable. We consider a Gaussian
distribution centered at $\pi$, with standard deviation $\pi/4$ and 
trimmed at the edges ($0$, $2\pi$). 
Using RESM we calculated the quantum van Trees information for
different values of $|\alpha|$. The results are depicted
in~\ref{fig:implaserfig}~(left) as red dots. The line is obtained
using~(\ref{eq:two_elements_povms:general}) with
$\rho(\xi) = |\phi(\xi)\rangle \langle \phi(\xi)|$ as an ansatz. The
figure suggest that the family of POVMS proposed is a good ansatz.


\subsection{Coherent plus thermal state}
\label{sec:thermal}

As a final example, we consider estimating a parameter, chosen from 
a given distribution, encoded in a non-pure
state. In general, there are no analytical expressions for the quantum Fisher 
information for this case. 
We calculate $\widetilde{Z}_Q$ in order to bound the error 
in estimating the parameter.
We build upon the last example, considering a mixture 
of \eref{eq:rotated:coherent} and a thermal state. Let 
\begin{equation}\label{eq:mixture_coherent}
    \varrho(\theta) = e^{i\theta\hat{n}}
        \left[
            \epsilon|\alpha\rangle\langle\alpha| 
            + (1-\epsilon) \sum_n\frac{\langle n\rangle^n}
               {(1+\langle n\rangle)^{n+1}}|n\rangle\langle n|
        \right]e^{-i\theta\hat{n}}
\end{equation}
with
\begin{equation}
    \langle n\rangle
    =\left[\exp\left(\frac{\hbar\nu}{k_BT}\right)-1\right]^{-1}=|\alpha|^2\,
    \end{equation}
be the state in which the parameter ($\theta$) is encoded.
For the right side of fig. \eref{fig:implaserfig}, we used a Gamma distribution of the form,
\begin{equation}
    \gamma_{\alpha,\beta}(x) = \frac{e^{\frac{-x}{\beta}}x^{\alpha-1}\beta^{-\alpha}}{\Gamma(\alpha)}
\end{equation}
with $\alpha=4$ and $\beta=1.5$.


%

In \fref{fig:implaserfig}~(right) we show the numerical calculations
of $\widetilde{Z}_Q$ using the algorithm RESM. We compare it with the
case of an initial coherent pure state and with the ansatz
composed of the two outcome POVM \eref{eq:two_elements_povms:general}
with $\rho(\xi)=|\phi(\xi)\rangle\langle\phi(\xi)|$, see \eref{eq:mixture_coherent}.
As expected, $\widetilde{Z}_Q$ is larger for a pure state: a coherent
pure state is better for estimating a phase than the
mixture~(\ref{eq:mixture_coherent}).
\subsection{A note for reproducing results} 
The code implementation can be obtained in the repository 
\cite{repositorymetrology}.  To reproduce the results presented in
\sref{sec:qubit}, set the flag {\tt -o} to {\tt Qubit} and vary the flag {\tt
--EtaAngle} from 0 to $\pi$.
For the results in sections 
\ref{sec:phase:estimation} and
\ref{sec:thermal}, set the flag {\tt -o} to {\tt  CohPlusTher} and to {\tt CohPlusTherGamma} respectively. We also set
the temperature with {\tt  -T 0.001}, the mixing constant {\tt --MixConstant 0.5},
the number of times to sample the space with {\tt -s 150},
the dimension of the Hilbert space to describe the system with 
{\tt  --HilbertDim 7} and the number of outcomes of the POVM with {\tt
--Outcomedim 10}. For pure state, as in \sref{sec:phase:estimation},
set {\tt --MixConstant  1}.
The squared norm of $\alpha$ is set with the option {\tt --MeanPhotonNumb},
which can be varied to reproduce 
the plots. The whole data set can be obtained with the command {\tt make all}.
\section{Conclusions}
\label{sec:conclusions}

The random extreme sampling method (RESM) can be used to find
efficiently the maximum of a cost function over all possible quantum
measurements. Particularly it is useful to find limits in the
precision of parameter estimation, through the cost function known as
the quantum Fisher information, when the state to be measured is a
mixed state. It can also be used to find the optimal measurement
strategy by a given convex cost function by finding the POVM that
maximizes it, at a considerable lower computational cost. 

\section{Acknowledgements}
Support by  PASPA-DGAPA, UNAM, projects CONACyT 285754 and UNAM-PAPIIT IG100518, IN-107414. 
\section*{References}
\bibliography{bibliografia}
\end{document}